# Automatic Handgun Detection in X-ray Images using Bag of Words Model with Selective Search

D. Castro Piñol and E. Marañón Reyes

*Abstract*— **Baggage inspection systems using X-ray screening are crucial for security. Only 90% of threat objects are recognized from the X-ray system based in human inspection. Manual detection requires high concentration due to the images complexity and the challenges objects points of view. An algorithm based on Bag of Visual Word (BoVW) with Selective Search is proposed in this paper for handguns detection in single energy X-ray images from the public GDXray database. This approach is an adaptation of BoVW for X-ray baggage images context. In order to evaluate the proposed method the algorithm effectiveness recognition was tested on all bounding boxes returned by selective search algorithm in 200 images. The most relevant result is the precision and true positive rate (PPV = 80%, TPR= 92%). This approach achieves good performance for handgun recognition. In addition, it is the first time the Selective Search localization algorithm was tested in baggage X-ray images and showed possibilities with Bag of Visual Words.**

*Index Terms*— **Single energy baggage X-ray images, Bag of Visual Words, Selective Search, computer vision**

## I. INTRODUCCIÓN

Desde que los rayos X fueron descubiertos en 1895 han sido muy utilizados para identificar estructuras internas de diversos objetos. No solo se han utilizado en aplicaciones médicas sino también en pruebas no destructivas (*Non-destrutive testing*). Las pruebas no destructivas son también conocidas como pruebas de rayos X (*X-ray Testing* del inglés) que son utilizadas para determinar si un objeto cumple con un conjunto de especificaciones necesarias [1].

Existen diferentes aplicaciones de pruebas de rayos X de las cuales muchas se han automatizado y semi-automatizado debido al uso de algoritmos de visión por computador.

Ejemplo de estas aplicaciones son la inspección de fallas en soldaduras, en cargamentos e inspección de la calidad en alimentos [2]. Uno de los campos de investigación que más ha llamado la atención en los últimos años es la automatización del proceso de inspección de equipajes en los aeropuertos y en puntos de seguridad.

La inspección de equipajes ha sido una importante aplicación de seguridad que ha llamado más la atención especialmente después del 11 de septiembre. Los sistemas de inspección de equipajes son usados para ayudar al personal de inspección a identificar objetos peligrosos. Sin embargo, el proceso de inspección tiene diversas dificultades. Estas dificultades están presentes en la complejidad que tiene identificar objetos peligrosos en este tipo de imágenes y en las circunstancias adversas que enfrentan los inspectores.

En estas imágenes, a diferencia de las imágenes de espectro visible, los píxeles representan el nivel de absorción del material frente al paso de los rayos X. Los rayos X son atenuados siguiendo la ley de absorción y densidad de la estructura de los objetos. Brinda información sobre la densidad y el material de los objetos. Los materiales metálicos suelen atenuar los rayos X más que otros materiales. A su vez, las imágenes de rayos X pueden ser de una sola energía (en escala de grises) o de doble energía (pseudocolor) en dependencia de la tecnología del equipo de adquisición [2].

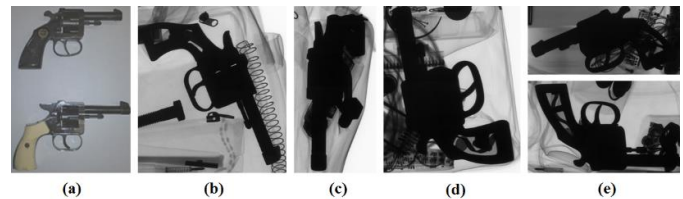

Figura 1. (a) Imágenes de revólveres en espectro visible tomada de [3] (b) revólver en imagen de rayos X, (c) Oclusión propia, (d) Oclusión con otros objetos, (e) otras posibles poses de los revólveres.

Los objetos peligrosos presentan complejidad en su identificación visual. Los objetos del equipaje en las imágenes de rayos X se pueden encontrar rotados, superpuestos unos con otros, con oclusión propia y con desafiantes puntos de vista. Además, estas imágenes tienen poca textura, son de bajo contraste, y los objetos suelen aparecer desordenados. También puede presentarse una adquisición errónea de la imagen y la presencia del ruido. En la figura 1 se evidencian algunas de estas características mencionadas.

La inspección de equipajes por imágenes de rayos X suele ser una tarea estresante para los inspectores debido a la gran cantidad de imágenes que deben procesar en un corto período de tiempo con elevada variabilidad de objetos y poses. Estos objetos tienen diferentes formas, texturas y sustancias como materiales metálicos, orgánicos e inorgánicos. La probabilidad humana de cometer errores se incrementa en el tiempo y con la diversidad de equipajes. Se ha registrado en [4] que la precisión manual en estos sistemas de inspección varía entre un 80 y 90% de detección de objetos peligrosos. Debido a que son muy pocas imágenes de equipajes que contienen objetos ilícitos en comparación con las que no presentan, estos procesos tienen una elevada razón de falsas alarmas. Estas razones pueden llevar también a que el proceso de inspección sea lento. Todo esto muestra la necesidad de automatización del proceso de inspección.

La automatización del proceso de inspección utilizando algoritmos de visión por computador permanece todavía como un problema no resuelto debido a las dificultades que presenta. Todavía se continúa realizando la inspección manual. En el trabajo de Mery [5] se explican cuatro causas generales que explican la dificultad que contiene la automatización de este proceso dentro del campo de visión computacional. Estas cuatro causas son: i) pérdida de generalidad, que significa que un método que funcione bien para una tarea puede no transferirse bien hacia otra, ii) la deficiente exactitud de la detección, que consiste en el compromiso entre las falsas alarmas y las detecciones que se pierden, iii) una robustez limitada y iv) una baja adaptabilidad.

En los años recientes ha habido diferentes trabajos investigativos dentro del campo de visión computacional y procesamiento digital de imágenes que buscan automatizar el proceso de inspección. Se han destacado dos líneas principales enfocadas al reconocimiento de objetos en imágenes de una sola energía y de doble energía. Algunos de los métodos que más se han destacado en la detección de objetos en imágenes de rayos X para ambos tipos de imágenes son BoVW [6] [7] [8], AISM [3], XASR [9], Visión Activa [10] y las Redes Neuronales Convolucionales (CNN) [11] [5]. También se han desarrollado investigaciones de reconocimiento de objetos utilizando imágenes de rayos X de múltiples vistas [12] [13] [14].

En este artículo se trabaja con el método Bolsa de Palabras Visuales BoVW para el reconocimiento y el algoritmo de Búsqueda Selectiva (*Selective Search*) [15] como detector. El método BoVW ha tenido resultados positivos en el reconocimiento de objetos en imágenes de espectro visible y en el reconocimiento de objetos en imágenes de rayos X de doble energía. A criterio de los autores se cree que el método de BoVW se ha probado poco en las imágenes de rayos X de una sola energía.

El objetivo de este trabajo es desarrollar y evaluar un algoritmo basado en el método de BoVW en las imágenes de equipaje de rayos X de una sola energía para el reconocimiento de armas (en este caso revólveres) utilizando como algoritmo de localización Búsqueda Selectiva. Para ello se utilizaron imágenes de revólveres presentes en la base de datos pública GDXray [16] y se utilizaron gran parte de los parámetros de BoVW configurados según los resultados de los experimentos en [17]. Las armas constituyen uno de los objetos más difíciles de detectar por su elevada variación intraclase debido a las diferentes poses que pueden adoptar y a su asimetría. En los trabajos presentados con esta base de datos se ha visto la dificultad que presenta el reconocimiento de este objeto frente a otros objetos [3].

La experimentación seguida consistió en utilizar los recuadros (*bounding boxes*) que retorna el algoritmo de Búsqueda Selectiva sobre las imágenes para realizar el entrenamiento, validación y prueba. Para el entrenamiento se utilizaron las imágenes de las serie B0049 y los recuadros de las imágenes de B0046, para la validación fueron los restantes recuadros de las imágenes de B0046. Finalmente para la etapa de prueba se utilizaron las imágenes de B0044.

Para el conocimiento de los autores se cree que es el primer trabajo que utiliza BoVW con el algoritmo de Búsqueda Selectiva en imágenes de equipaje de rayos X. Los aspectos más destacables de este artículo radican en la configuración utilizada de BoVW, en el uso combinado del algoritmo de Búsqueda Selectiva para la generación de recuadros en el protocolo de experimentación y la detección de las armas en las imágenes.

Este artículo está organizado de la siguiente forma: la sección II presenta una explicación del método propuesto, la sección III expone el protocolo de experimentación utilizado, la sección IV muestra los resultados obtenidos y la sección V las conclusiones del trabajo.

## II. METODOLOGÍA

En esta sección se explica la metodología usada donde se exponen las diferentes etapas y configuración de BoVW. Se expone la etapa de pre-procesamiento utilizada, el algoritmo de extracción de características, el vocabulario de palabras visuales construido, el algoritmo Búsqueda Selectiva y el clasificador utilizado. También se presentan los valores de los parámetros seleccionados utilizados en la implementación.

El método Bolsa de Palabras Visuales (*Bag of visual words*) [18] [19] es muy conocido y se ha utilizado para la clasificación de imágenes por contenido y el reconocimiento de objetos en imágenes. Esta metodología fue primero propuesta como Bolsa de Palabras (*bag of words*) para la clasificación de textos por contenido donde el vector de características es formado por la frecuencia de aparición de palabras clave. Luego el modelo de bolsa de palabras fue extendido para aplicaciones de visión de computador donde se le conoce también como Bolsa de Palabras Visuales. Una de las principales ventajas de BoVW es su simplicidad, su eficiencia computacional y su cierta invariancia a transformaciones como la oclusión y la iluminación según lo expuesto en [18]. Esta metodología está orientada a construir el vector de rasgos y está compuesta por diversos pasos.

La metodología de BoVW está compuesta por los siguientes pasos: (i) detección automática de puntos de interés en un conjunto de imágenes, (ii) cálculo de los descriptores clave sobre estos puntos de interés, (ii) agrupamiento de los descriptores en diferentes grupos llamados "palabras visuales" para formar el vocabulario de palabras visuales y (iv) dada una nueva imagen formar una representación basada en la ocurrencia de las palabras visuales del vocabulario encontradas en la imagen. Esta representación es un histograma de los rasgos visuales de la imagen [20].

El modelo Bolsa de Palabras Visuales se define también de la siguiente forma. Dado un conjunto de datos de entrenamiento $D$ que tiene $n$ imágenes representado por $D = d_1, d_2,..., d_n$ donde $d$ son el conjunto de descriptores de las características visuales de cada imagen. Luego un algoritmo de aprendizaje no supervisado, como el $k$-means, es utilizado para agrupar $D$ en un conjunto de palabras visuales $W$ representado por $W = w_1, w_2,..., w_v$ donde $V$ es la cantidad de grupos. Los datos se resumen entonces en una tabla de $V \times N$ ocurrencias de $N_{ij} = n(w_i, d_j)$ donde $n(w_i, d_j)$ expresa cuántas veces la palabra visual $w_i$ ocurre en una imagen $d_j$

El primero de los pasos del método BoVW es la detección automática de puntos de interés visual en las imágenes. Estos puntos de interés o puntos clave son puntos con coordenadas $x$ y $y$ en la imagen obtenidos con un algoritmo detector que representan áreas con información visual relevante. Son comunes los detectores de SIFT [21], SURF [22] y Harris [23]. Por otro lado los descriptores describen la información de estos puntos. Como por ejemplo los descriptores SIFT [21] que son un histograma espacial de los gradientes de las regiones de los puntos de interés, constituyen un vector de 128 dimensiones. El éxito de los descriptores SIFT radica en su parcial invarianza a la rotación, a la escala y algunos cambios de iluminación.

Luego estos descriptores son agrupados mediante un algoritmo de aprendizaje no supervisado como el $k$-means. El propósito de $k$-means es dividir un conjunto de vectores en $k$ grupos distintos alrededor de un vector media común. Es decir, trata de encontrar $k$ centros conocidos también como *centroides*. Estos *centroides* deben representar al patrón compartido por los puntos claves en ese grupo. Estos grupos son las llamadas palabras visuales y una colección de estas palabras se conoce como vocabulario visual.

### A. Pre-procesamiento

Los materiales metálicos son los que más absorben los rayos X, razón por lo que en las imágenes de rayos X de una sola energía constituyen las áreas más oscuras. Se puede apreciar en las imágenes de la figura 1 que los revólveres presentan píxeles que van desde un gris oscuro hasta el negro representando el elevado nivel de absorción del material por los rayos X. Para aprovechar esta característica y no extraer información visual de poco interés se realiza un pre-procesamiento de la imagen antes de la extracción de los puntos clave. Este pre-procesamiento tiene como objetivo evitar extraer innecesaria información del fondo de la imagen de rayos X. Se ha comprobado en algunos trabajos [14] [7] la eficiencia de esta técnica en la precisión del algoritmo.

Esta etapa de pre-procesado consiste principalmente en una segmentación utilizando un umbral de nivel de gris determinado experimentalmente para formar una imagen binaria [24]. Esta experimentación consistió en la búsqueda de un valor de nivel de gris que cubriera los valores de grises de los objetos metálicos. El umbral seleccionado es 0,31 para una imagen de rayos X de 8 bits normalizada, donde el valor 1 corresponde al nivel 255.

Luego, con la imagen binaria obtenida se rechazan las regiones de convergencia que tienen un área menor que la décima parte de las dimensiones de la imagen. Esta imagen binaria resultante es la máscara que se utiliza para extraer las características visuales de interés.

### B. Extracción de características

La imagen binaria obtenida del pre-procesamiento se utiliza como soporte para extraer las características visuales de las regiones resultantes. En este trabajo se utilizó el algoritmo PHOW (*Pyramid Histogram Of Visual Words*) para la extracción de características propuesto en [25]. El mismo calcula una cantidad de puntos definidos a priori a cuatro escalas fijas y está basado en los descriptores de SIFT. Se ha reportado el uso de PHOW para el reconocimiento de objetos con BoVW en imágenes de equipaje de rayos X [26] [17].

El algoritmo PHOW realiza un muestreo denso de puntos con espacio de $M$ píxeles donde en este trabajo se seleccionó $M = 4$. Los autores consideran que esta técnica es ventajosa por tratarse de imágenes con poca textura, por lo que es necesario una mayor extracción de características. Se ha reportado en la literatura las ventajas que presenta esta técnica [27]. Como también se ha reportado un incremento en la precisión de la clasificación a mayor cantidad de puntos extraídos [27] [28].

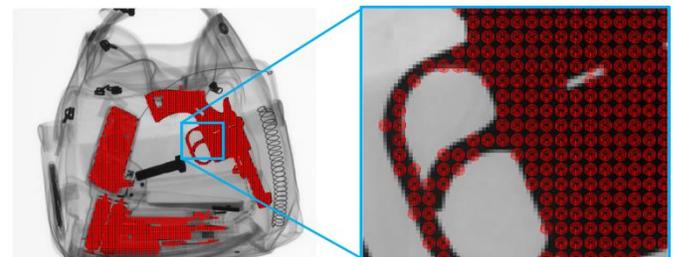

Figura 2. Puntos clave PHOW en una imagen de rayos X de equipaje.

Por otro lado, las escalas son definidas modificando el ancho de la ranura espacial de los descriptores SIFT a 4, 6, 8 y 10 píxeles respectivamente. Razón por la que cada punto clave es representado por cuatro descriptores SIFT. Estas escalas fueron usadas también para filtrar la imagen antes de obtener los descriptores por un filtro Gaussiano con desviación estándar $\sigma = SIZE / MAGNIF$, donde *SIZE* son las cuatro escalas mencionadas anteriormente y *MAGNIF = 6* es el factor de magnificación del descriptor. Además, este filtrado es útil para la eliminación de posible ruido presente en la imagen. Para mayor información sobre esta configuración revisar la implementación de PHOW en la biblioteca de funciones VLFeat [29]. En la figura 2 se puede apreciar un ejemplo de los puntos clave de PHOW.

## C. Vocabulario de palabras visuales

Las imágenes utilizadas para la generación del vocabulario de palabras visuales fueron de las series B0046 y B0044 de la base de datos GDXray. Se utilizó el clásico *k-means* como algoritmo de aprendizaje no supervisado. Se definió a priori el tamaño del vocabulario como $V = 1000$ que resulta un valor adecuado que se ha usado en otros trabajos. Para la solución se realizaron diez ejecuciones del algoritmo y se seleccionó el vocabulario resultante con menor función de costo. Debido a la inicialización aleatoria del algoritmo *k-means* es que se realiza esta técnica.

Para la construcción del vocabulario visual se utilizaron todas las características visuales extraídas de los materiales metálicos de las imágenes. Se utilizaron los descriptores de las regiones metálicas de las imágenes como se puede apreciar en la figura 3. Esta técnica de utilizar las regiones metálicas mostró tener buenos resultados en el trabajo [17] y al vocabulario resultante se le llamó vocabulario metálico. El vocabulario metálico es un vocabulario adaptado que ofrece una mejor representación de los objetos metálicos que un vocabulario universal. Esto se debe a que los objetos metálicos están siendo representados en un vector de rasgos de mayor dimensión. Este razonamiento fue inferido gracias al trabajo de [30].

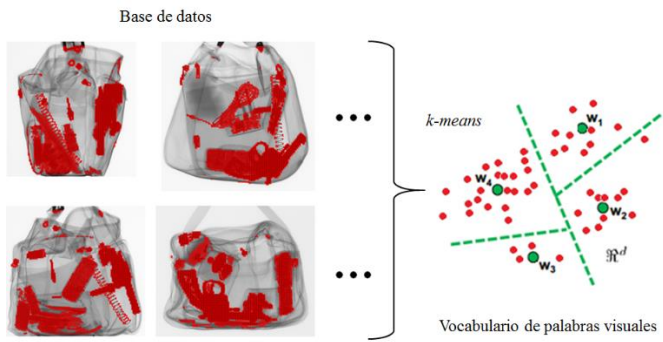

Figura 3. Agrupamiento de características mediante *k-means* y formación del vocabulario de palabras visuales. En este ejemplo $V = 4$

Destacar que el pre-procesamiento utilizado para la extracción de características utilizadas en la generación del vocabulario visual difiere de la explicada anteriormente en que no se realiza un rechazo de regiones de pequeñas áreas.

## D. Histogramas de palabras visuales

Después de obtener el vocabulario visual se pueden construir los histogramas de palabras visuales sobre nuevas imágenes. Estos vectores de rasgos fueron construidos utilizando el estándar de asignación dura (*hard assigment*) que consiste únicamente en la asignación de cada descriptor de la nueva imagen a la palabra visual más cercana del vocabulario. Luego los histogramas fueron normalizados mediante la expresión:

$$N'_{ij} = \frac{N_{ij}}{\sum_i N_{ij}} \qquad (1)$$

Donde $N_{ij}'$ es la cantidad de descriptores normalizados asociados a la palabra visual $w_i$ donde toma valores entre cero y uno. Independientemente de que no se hayan utilizado algunas técnicas conocidas [31] para la incorporación de la información espacial en los histogramas los mismos poseen cierta información espacial debido al solapamiento entre los descriptores [32], producto del uso de PHOW.

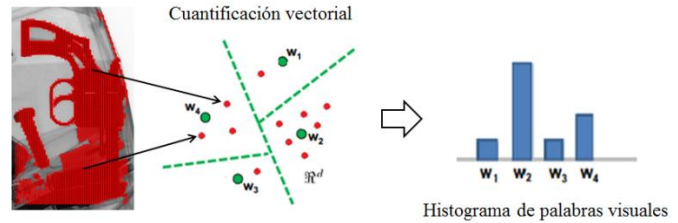

Figura 4. Cuantificación vectorial para formar el histograma de palabras visuales.

## E. Búsqueda Selectiva como algoritmo detector

La Búsqueda Selectiva es un algoritmo de detección de objetos en imágenes presentado en [15]. Este algoritmo aprovecha toda la estructura de la imagen porque está basado en la segmentación en diferentes niveles jerárquicos. Propone una diversificación de técnicas de muestreo para cubrir la mayor variedad de condiciones posibles en las imágenes.

Se destaca porque utiliza un agrupamiento jerárquico para lidiar con todas las posibles escalas de los objetos. Utiliza diversas estrategias de agrupamiento donde varía con: el espacio de color y funciones de similitud para lidiar con la diversa naturaleza de los objetos en cuanto a la textura, el tamaño y/o una medida de la incidencia.

Debido a la calidad de los recuadros hipótesis generados por el algoritmo de Búsqueda Selectiva sobre los objetos posibilita su uso con el método BoVW [15]. Este algoritmo genera un número significativamente reducido de recuadros de hipótesis de objetos en comparación con la búsqueda exhaustiva (método de la ventana deslizante). Un total de 2k-3k de recuadros hipótesis en comparación con 62500000k del método de ventana deslizante. También es reconocido por brindar una buena calidad en la detección de los objetos.

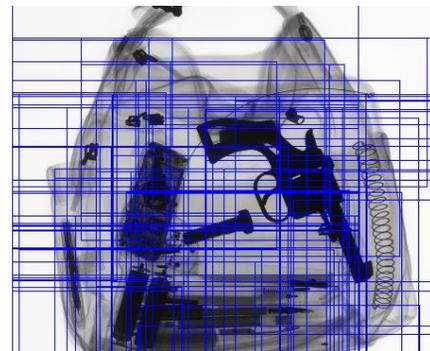

Figura 5. Ejemplo de los recuadros (seleccionados 200) que devuelve Búsqueda Selectiva en una imagen de rayos X de equipaje ($k = 100$ y $\sigma = 0.8$).

Entre los principales parámetros que se pueden modificar en el algoritmo está $\sigma$ que pertenece al filtro Gaussiano para eliminar posibles artefactos. También está el parámetro $k$ que controla el tamaño de los segmentos de la segmentación inicial. Este parámetro configura la escala de observación de manera que valores elevados de $k$ son preferentes para componentes más grandes y viceversa. Estos parámetros fueron presentados inicialmente en [33].

En la figura 5 se puede apreciar un ejemplo de los recuadros generados por el algoritmo en una imagen de rayos X de equipaje. Para esta imagen se generó un total 2093 recuadros de hipótesis.

Se ha manifestado que el algoritmo de Búsqueda Selectiva puede utilizarse con BoVW aunque los autores no han encontrado un ejemplo de su aplicación en imágenes de rayos X de equipaje. La aplicación de BoVW con el método de Búsqueda Selectiva consiste en la evaluación del clasificador entrenado en cada recuadro hipótesis que devuelve la Búsqueda Selectiva.

### F. Clasificador SVM

Para el reconocimiento de las armas en las imágenes de rayos X se utilizó un clasificador de Máquina de Soporte Vectorial SVM que es un algoritmo de aprendizaje supervisado muy conocido. Se ha mostrado en varios trabajos el éxito que presenta el uso de la SVM como clasificador con los vectores de rasgos de BoVW.

Una SVM construye un hiperplano o conjunto de hiperplanos en un espacio de dimensionalidad muy alta. Existen infinitos hiperplanos de separación. El objetivo de la SVM está en encontrar el hiperplano que separe dos conjuntos finitos de datos con el máximo margen posible [34]. El margen es definido como la distancia del punto más cercano de entrenamiento al hiperplano de separación.

$$h(\boldsymbol{x}) = \boldsymbol{w}^T \boldsymbol{x} + b \qquad (2)$$

$$E(\boldsymbol{w}) = \frac{\lambda}{2}\|\boldsymbol{w}\|^2 + \frac{1}{n}\sum_{i=1}^{n}\ell_i(\langle \boldsymbol{w}, \boldsymbol{x}\rangle) \qquad (3)$$

$$\ell_i(\langle \boldsymbol{w}, \boldsymbol{x}\rangle) = (y_i - \langle \boldsymbol{w}, \boldsymbol{x}\rangle)^2 \qquad (4)$$

Se puede apreciar en (2) la hipótesis $h(\boldsymbol{x})$ de la SVM o función de decisión donde $\boldsymbol{w}$ y $b$ son los parámetros del hiperplano de máximo margen. Si $h(\boldsymbol{x}) > 0$ entonces el dato $\boldsymbol{x}$ pertenece a la clase y si no viceversa. En este ejemplo, cero es el umbral de decisión.

En (3) $E(\boldsymbol{w})$ es la función objetivo regularizada o función de costo usada en la implementación de este trabajo. El objetivo radica en encontrar el valor de $\boldsymbol{w}$ que minimice la función de costo dado los datos $\boldsymbol{x}$ de entrenamiento y las etiquetas $y_i$ para cada dato. También se encuentra $\lambda$ que es el parámetro de regularización del algoritmo de entrenamiento. El mismo presenta una relación inversamente proporcional con el conocido parámetro $C$ del entrenamiento de las SVM. En la sección III $E$ se explica el proceso de selección del parámetro de regularización del algoritmo. Esta función objetivo fue tomada de VLFeat [35]. En la (4) se aprecia la función de pérdida (loss function) $\ell_i$ que expresa una medida del error de los datos $x_i$ con sus etiquetas $y_i$. La función de pérdida en (4) fue la que se utilizó en este trabajo y se le conoce también como L2. Fue seleccionada a que presentó los mejores resultados en los experimentos que se realizaron en [17].

Los datos no son siempre linealmente separables. Las SVM resuelven esto mediante dos formas: i) expandiendo los datos a una mayor dimensión donde tengan una mayor separación lineal mediante la función *kernel* y ii) permitir muestras mal clasificadas como parte de la solución y penalizarlas en proporción a su distancia a la frontera de decisión.

Debido a que las clases generalmente no son linealmente separables, se hace necesario entonces expandir los datos a una dimensión superior a través de un *kernel* para poder encontrar una mayor separación lineal de las clases. Mientras los *kernels* lineales son muy eficientes en el entrenamiento, los *kernels* no lineales tienden a brindar mejores resultados en la precisión de la clasificación [28]. Una clase de *kernels* que son casi tan eficientes como los lineales pero usualmente con mayor precisión son los *kernels* homogéneos aditivos [36]. Este tipo de *kernel* fue el utilizado en este trabajo.

El *kernel* homogéneo aditivo utilizado fue el $\chi^2$ que se ha reportado su buen desempeño en los trabajos [17] [37] [36]. Este kernel homogéneo aditivo fue configurado con un orden de expansión igual a 2 permitiendo que los histogramas tuvieran una dimensión cinco veces mayor. El grado de homogeneidad del *kernel* seleccionado fue de $\gamma = 1$, de manera que el *kernel* es homogéneo puro.

### G. Materiales usados en la implementación

Este trabajo se realizó en el entorno de desarrollo de MATLAB 2015b donde se programaron los experimentos realizados. Se utilizó la biblioteca de funciones de código abierto VLFeat v0.9.2 para el algoritmo *k-means*, para la extracción de características con PHOW y los descriptores de SIFT. Las implementaciones del clasificador SVM, dígase el entrenamiento y la expansión de los mapas homogéneos aditivos fueron tomadas también de VLFeat. Para la detección de las armas se utilizó la implementación del algoritmo de Búsqueda Selectiva publicada para MATLAB en [15].

En la mayoría de los experimentos, principalmente las etapas off-line de entrenamiento, se utilizó computación de alto rendimiento HPC[2] (*High Performance Computing*) principalmente para la extracción de características de las imágenes que se usaron con sus dimensiones originales, para la ejecución de *k-means* y el algoritmo de Búsqueda Selectiva. La gran mayoría de experimentos de evaluación fueron ejecutados en este HPC.

## III. Experimentación

En esta sección se presenta la metodología de experimentación que se utilizó y los experimentos realizados para evaluar la precisión en el reconocimiento y detección de las armas en estas imágenes de rayos X.

Todas las imágenes usadas en los experimentos de este

---

trabajo fueron tomadas de la base de datos GDXray[3] [16]. Esta base de datos es de carácter público y contiene más de 20 000 imágenes de diferentes aplicaciones de pruebas de rayos X, es la más usada en la literatura que utilizan imágenes de rayos X de equipaje de una sola energía.

Las imágenes de armas de equipaje en GDXray son desafiantes para un algoritmo de reconocimiento debido a su elevada variación intraclase. Algunos ejemplos aparecen en la figura 1. Las series utilizadas en los experimentos fueron la B0044, B0046 y B0049. Las mismas tienen 178, 200 y 200 imágenes respectivamente. Las dos primeras series son imágenes de equipajes que contienen una o dos armas y tienen unas dimensiones de $2208 \times 2688$ píxeles. La última está compuesta únicamente por imágenes de revólveres en diferentes posiciones y puntos de vista con unas dimensiones alrededor de $626 \times 557$ píxeles para propósitos de entrenamiento. Todos los experimentos realizados fueron hechos con las dimensiones originales de las imágenes.

Las armas presentes en las bases B0044 y B0046 fueron anotadas con recuadros delimitadores debido a que las anotaciones presentes en GDXray son recuadros con orientación y el algoritmo de Búsqueda Selectiva los devuelve sin orientación. Para ello se utilizó la herramienta X-vis presentada en [2] que implementa la función `Xannotate` para realizar estas anotaciones.

### A. Evaluación de la detección de Búsqueda Selectiva

Lo primero que se realizó fue una evaluación del algoritmo de detección de Búsqueda Selectiva para las armas presentes en las series B0044 y B0046 para buscar el parámetro $k$ que tuviera mejor precisión en la detección de armas. Esta evaluación se realiza porque el resultado obtenido sería la cota superior máxima en la calidad de la detección de las armas.

Esta prueba tiene gran importancia porque el algoritmo de Búsqueda Selectiva se ha probado principalmente en imágenes de espectro visible según la literatura. De manera que constituye esta evaluación una primera aproximación de la calidad de la detección de objetos en imágenes de rayos X de equipaje usando este algoritmo.

La métrica utilizada para evaluar la precisión del algoritmo de Búsqueda Selectiva fue el criterio de superposición entre el área rectangular de la detección y el del área rectangular etiquetada en PASCAL *Visual Object Classes Challenge* [38]. Este criterio $\alpha_0$ es la razón entre el área de intersección sobre el área de la unión y es considerada correcta una detección cuando $\alpha_0 \geq 0.5$. En (5) se aprecia la definición de $\alpha_0$

$$\alpha_0 = \frac{area(BB_{dt} \cap BB_{gt})}{area(BB_{dt} \cup BB_{gt})} \qquad (5)$$

Donde $BB_{dt}$ es el recuadro delimitador (*Bounding Box*) de la detección que devuelve el algoritmo de Búsqueda Selectiva y $BB_{gt}$ es el recuadro delimitador anotado del arma. Luego se utiliza la métrica de la razón de aciertos positivos. Esta se calcula mediante la razón de recuadros delimitadores que obtuvieron un criterio de solapamiento superior al umbral de Pascal por cada imagen con armas sobre el total de las imágenes. Brinda una medida de la calidad de la detección.

El algoritmo se configuró para que utilizara todas las funciones de similitud y para escala de grises. Fue utilizado un valor de $\sigma = 0.8$. Se realizaron diversas ejecuciones del algoritmo modificando el valor de $k$ desde 100 hasta 900. Se pudo apreciar que a mayor $k$ menor la precisión de la detección. Se seleccionó $k = 100$ que devuelve una razón de 84% de detección sobre todas las armas. Este es un resultado bajo de precisión razón por lo que se experimentó con los umbrales de Pascal de [0.4 0.45 0.5]. Se decidió utilizar un umbral de 0.4 que brinda un resultado de 98% en la detección de las armas. Algunos autores debido a las características y circunstancias de los objetos a detectar varían el umbral de Pascal hasta 0.4 como se realizó en [3]. Esta primera evaluación de la calidad de la localización del algoritmo del Búsqueda Selectiva tomando como referencia las anotaciones de las armas, constituye una contribución experimental en este contexto de aplicación.

Otro cálculo que se realizó fue el promedio de las mejores detecciones. Es decir, se tomaron los valores máximos de $\alpha_0$ en cada imagen con arma y se realizó un promediado donde se obtuvo un valor final de 0.618. Este indicador es una medida promedio de las mejores detecciones sobre todas las imágenes con armas que se encuentre por encima del umbral de Pascal 0.5.

### B. Generación de las imágenes de entrenamiento y prueba

La metodología propuesta para conformar los grupos de entrenamiento, validación y prueba se basó en el algoritmo de Búsqueda Selectiva. Es decir, se utilizaron las ventanas que devuelve el algoritmo de Búsqueda Selectiva sobre las imágenes de equipaje de las series B0044 y B0046 para generar los grupos de entrenamiento, validación y prueba. Además, se utilizaron las imágenes positivas de armas de B0049 para el entrenamiento. En la tabla I se presenta la distribución de las imágenes usadas y la cantidad de ventanas para cada grupo. Señalar que los grupos están balanceados sobre la cantidad de imágenes positivas y negativas.

Para delimitar los grupos positivos y negativos las ventanas que tuvieron un índice de solapamiento mayor a 0.4 con respecto a las anotaciones de cada arma son etiquetadas como imágenes positivas y las restantes como negativas.

TABLA I.
DISTRIBUCIÓN DE LAS IMÁGENES DE GDXRAY
EN LOS GRUPOS DE EVALUACIÓN

| Imágenes | Series | Cantidad |
|---|---|---|
| Entrenamiento | B0049 | 200 |
| | B0046 | 25248 |
| Validación | B0046 | 10820 |
| Prueba | B0044 | 31894 |

### C. Métricas de evaluación

Para la evaluación del desempeño del clasificador en el reconocimiento de armas se utilizaron las métricas: razón de verdaderos positivos (VPR) o sensibilidad y la precisión (PPV) o valor predictivo positivo. En (6) y (7) se pueden apreciar respectivamente estas métricas.

---

[3]La base de datos GDXray es de dominio público y se encuentra en: http://dmery.ing.puc.cl/index.php/material/gdxray/

$$VPR = \frac{VP}{VP + FN} \tag{6}$$

$$PPV = \frac{VP}{VP + FP} \tag{7}$$

$$F1 = \frac{2 * VPR * PPV}{VPR + PPV} \tag{8}$$

Donde los verdaderos positivos (VP) son las armas correctamente clasificadas, los falsos positivos (FP) son los objetos no armas clasificados como armas, también se le conoce como falsas alarmas o error de tipo I. Los falsos negativos (FN) que son el número de armas clasificadas como no armas conocido también como error de tipo II. Por otro lado, se puede apreciar la métrica F1 en (8) que es una medida que depende de PPV y VPR muy conocida para realizar comparaciones entre diferentes pares de PPV y VPR.

### D. Protocolo de experimentación

El protocolo de experimentación que se siguió en este trabajo para evaluar el reconocimiento de las armas por el clasificador SVM está constituido por una etapa de entrenamiento, de validación y de prueba. En cada etapa se utiliza la distribución presentada en la tabla I. En este experimento de entrenamiento y evaluación se utilizó el método de retención o *holdout* ya que los datos fueron separados en estos tres conjuntos mutuamente excluyentes. Primero se realizaron un total de 10 corridas de selección aleatoria de los grupos de experimentación, donde se pudo observar cómo los resultados no presentaron variación (0.01%) significativa debido a la elevada cantidad de imágenes representativas usadas. Esta es la razón por la que en este estudio no se consideró el uso de la validación cruzada.

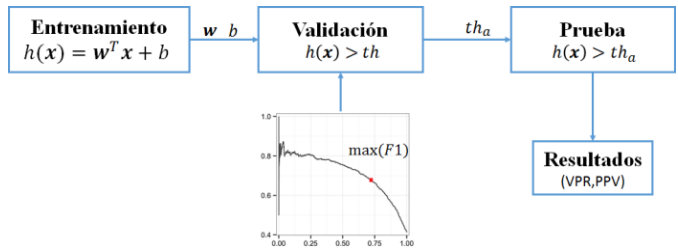

Figura 6. Protocolo de experimentación utilizado

La etapa de entrenamiento del clasificador SVM obtiene como resultado el vector $\boldsymbol{w}$ y el bias $b$ que son los parámetros del hiperplano de máximo margen que separa ambas clases. Luego, en la etapa de validación se realiza una variación del umbral de decisión $th$ de manera que se asume como positiva una muestra cuando la hipótesis de la SVM sea mayor que el umbral $h(\boldsymbol{x}) > th$ y en caso contrario viceversa. De esta variación se obtiene como resultado una curva *Precision-Recall* donde el eje $x$ está compuesto por los valores de VPR (*Recall*) y el eje $y$ por los valores de PPV (*Precision*). De esta curva se seleccionó el punto que tiene el máximo F1 de (8) que devuelve el umbral de decisión ajustado $th_a$ Después con este valor de umbral se procede a evaluar finalmente el grupo de prueba donde se obtiene un único par de PPV y VPR que constituyen el resultado final a presentar del trabajo. En la figura 6 se presenta la secuencia del protocolo de experimentación explicada.

### E. Ajuste del parámetro regularizador

Los parámetros del entrenamiento del clasificador fueron fijados a los valores escogidos que se expusieron en la sección II que coinciden con los resultados de los experimentos presentes en [17]. El principal parámetro que se ajusta en todo el protocolo de experimentación es el parámetro regularizador $\lambda$. El procedimiento de selección de este parámetro tiene gran significancia para evitar que ocurra un sobre-ajuste (*overfitting*) o sub-ajuste (*underfitting*) sobre el algoritmo de entrenamiento.

El sobre-ajuste de un algoritmo clasificador ocurre cuando los parámetros del algoritmo son ajustados de tal manera que aprende a reconocer muy bien los datos de entrenamiento pero pierde capacidad de generalización para el reconocimiento de nuevos datos. Sin embargo el sub-ajuste se presenta cuando ha sido un pobre aprendizaje y no se reconoce correctamente los datos de entrenamiento y los de prueba.

La mejor selección del parámetro regulador $\lambda$ se realiza mediante las curvas de aprendizaje. Una curva de aprendizaje es una representación gráfica del desempeño del clasificador para cada grupo de experimentación contra la variación del parámetro regularizador $\lambda$ o también mediante la variación en la cantidad datos.

En este trabajo se presenta la curva de aprendizaje del desempeño de los grupos de validación y prueba medidos con una métrica de error contra diferentes valores de $\lambda$. En este apartado se define el error como el promedio de predicciones incorrectas. En la figura 7 se puede apreciar las curvas de aprendizaje obtenida de estos experimentos.

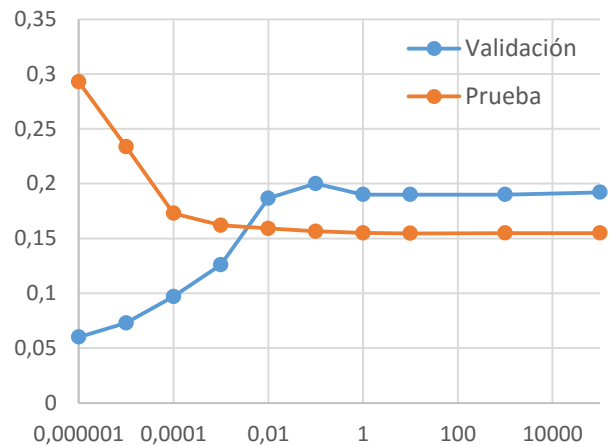

Figura 7. Curvas de aprendizaje de la validación y la prueba. El eje $y$ es el error y el eje $x$ es $\lambda$

Se puede apreciar en la figura 7 como las curvas de aprendizaje muestran la existencia de sobre-ajuste en el algoritmo para valores muy cercanos a cero del parámetro regularizador $\lambda$. Este sobre-ajuste se aprecia en la marcada diferencia del error del grupo de validación y de prueba para valores de $\lambda$ menores de 0.01 donde esa diferencia aumenta a medida que $\lambda$ se acerca a cero. Señalar que la validación, aunque se realiza sobre datos no utilizados en el entrenamiento de cierta forma también forma parte de un "entrenamiento" puesto que se realiza un ajuste del umbral de decisión. Finalmente se seleccionó $\lambda = 10$ ya que aproximadamente a partir de este valor comienza a apreciarse estabilidad en el error para ambos grupos. Esto muestra que para este valor se logra la generalización del clasificador en el reconocimiento de datos no vistos.

## IV. Resultados y Discusión

En esta sección se exponen los resultados finales de este trabajo utilizado el método expuesto en la sección II y el protocolo de experimentación presentado en la sección III para medir el desempeño del reconocimiento del clasificador. Finalmente se presentan los resultados del desempeño del detector.

### A. Resultados en el reconocimiento del clasificador

Los resultados del desempeño en el reconocimiento del clasificador se pueden apreciar en la tabla II.

TABLA II.
Resultados Obtenidos en el Protocolo de Experimentación.

| Métricas | Validación | Prueba |
|----------|-----------|--------|
| PPV | 74% | 80% |
| VPR | 94% | 92% |
| F1 | 83% | 85% |
| Error | 0.19 | 0.1546 |

Los resultados expuestos en la validación son los correspondientes al punto de máximo F1 seleccionado de la curva *Precision-Recall* que se puede apreciar en la figura 8. Finalmente los resultados obtenidos en el conjunto de prueba son PPV = 80% y VPR = 92% como métricas del desempeño del reconocimiento del clasificador.

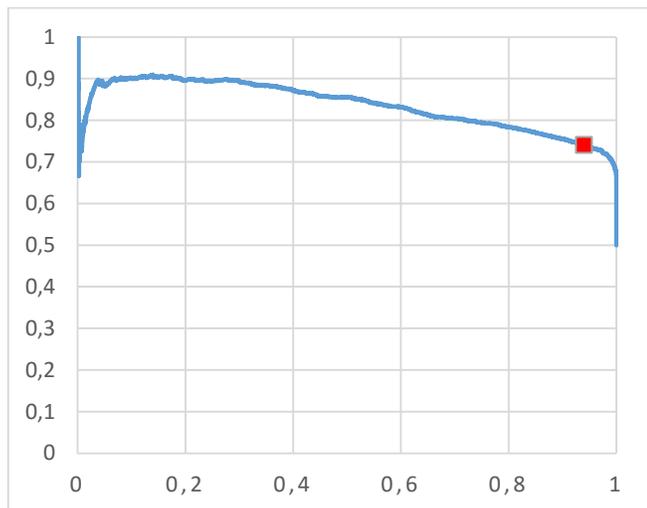

Figura 8. Curva *Precision-Recall* obtenida en la validación. Eje $x$ corresponde a los valores de VPR y eje $y$ los valores PPV.

Los autores consideran que en el trabajo presentado en [17] hubo sobreajuste en el entrenamiento del clasificador debido al valor cercano a cero del parámetro regulador ($\lambda = 0.00001$) y de que no se realizó una etapa final de prueba para verificar la no existencia de sobreajuste. De manera que es significativo el uso de las curvas de aprendizaje en este contexto para encontrar un valor de $\lambda$ adecuado.

### B. Resultados del desempeño del detector

Luego de obtener los resultados del desempeño del reconocimiento del clasificador se procede a exponer los resultados de la calidad de la detección.

El procedimiento final para detectar las armas online consiste en: i) realizar la segmentación en la imagen para utilizar las regiones metálicas, ii) extracción de las características estas regiones, iii) obtención de los recuadros hipótesis del algoritmo de Búsqueda Selectiva, iv) cálculo de los histogramas de palabras visuales en cada recuadro hipótesis y rechazar los recuadros que obtuvieron clasificación negativa con la SVM, v) del conjunto de recuadros positivos remover los *outliers*, vi) finalmente fusionar (*merge*) los recuadros restantes para obtener una única detección.

En la figura 9 se pueden apreciar algunos ejemplos de detección de armas de la serie B0044 con sus respectivos índices de solapamiento.

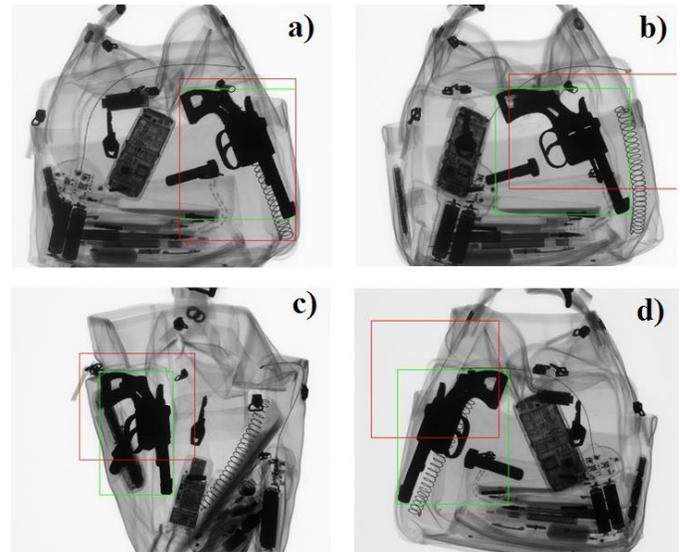

Figura 9. Ejemplo de detecciones finales en la serie B0044. Los recuadros verdes son las anotaciones y los recuadros rojos la detección final Los índices de solapamiento fueron: a) $\alpha_0 = 0.8$ b) $\alpha_0 = 0.41$ c) $\alpha_0 = 0.43$, d) $\alpha_0 = 0.3$

Finalmente el resultado obtenido en la detección fue de un 37,5 % del total de armas de B0044 fueron detectadas con un índice de solapamiento mayor o igual que 0.4. Esto representa un total de 67 imágenes con armas localizadas del total de 178 imágenes.

Para evitar confusión con este resultado hay que tener en cuenta la diferencia entre el reconocimiento y la detección ya que en algunas literaturas ambos términos se mencionan indistintamente. En este trabajo, el reconocimiento es la capacidad del algoritmo de reconocer la presencia de las armas en las imágenes y la detección es la capacidad de localizar la posición del arma en las imágenes originales de equipaje.

### C. Análisis y discusión

En este apartado se presenta una comparación y discusión de los resultados obtenidos con algunos trabajos similares reportados en la literatura que utilizan BoVW. Primero, se presentan algunos comentarios sobre los parámetros del método BoVW, a modo de comparación, con los trabajos de Bastan [6] y Turcsany [7]. También se realiza una comparación metodológica y experimental con una de las propuestas del reciente trabajo de Mery [5] que utiliza BoVW en una de las técnicas presentadas con la misma base de datos de este trabajo. La tabla III agrupa los principales parámetros de BoVW usados por cada trabajo y sus resultados.

Es importante destacar que los estudios de Bastan [6] y Turcsany [7] utilizan diferentes bases de datos y diferente tipo de imagen (doble- energía) razón por la que una comparación con ellos no debe ser estrictamente basada en los resultados, específicamente cuando los resultados son cercanos. De manera general, se puede apreciar en la tabla III como los resultados obtenidos del presente estudio son competitivos en PPV y VPR frente a demás trabajos.

Las principales diferencias de este trabajo con los trabajos de Bastan [6] y Turcsany [7] se concentran en la extracción de características, el tipo de vocabulario visual construido y el tipo de kernel usado en la SVM.

TABLA III.
COMPARACIÓN DE ALGUNOS PARÁMETROS DE BOVW CON OTROS TRABAJOS SIMILARES.

| Parámetro | Bastan [6] | Turcsany [7] | Mery [5] | Actual estudio |
|---|---|---|---|---|
| Detección de puntos | DoG+Harris | SURF | SIFT | PHOW |
| Descriptor | SIFT | SURF | SIFT +LBP[4] | SIFT |
| Vocabulario visual | Universal 200[5] | Formado por dos clases 1200 | dos vocabularios concatenados 300 | Metálico 1000 |
| Clasificador (kernel) | SVM (Intersección de Histogramas) | SVM (RBF) | Random forest | SVM ($\chi^2$) |
| PPV | 29% | *[6] | 65% | 80% |
| VPR | 70% | 99,1% | 84% | 92% |

En esta propuesta se usó como extracción de características PHOW que se considera relevante debido a la técnica de muestreo denso de puntos y su ventaja para este tipo de imágenes de poca textura. Los autores lo consideran más apropiado frente a otros detectores como SIFT y SURF para esta aplicación por su mayor cantidad de puntos extraídos y velocidad. En cuanto a los descriptores utilizados en los trabajos se ha destacado el descriptor SIFT que continua presentando elevado uso en aplicaciones de visión por computador. No obstante los autores consideran que quizás el descriptor usado por Mery ofrezca mejor caracterización de los puntos ya que está compuesto por descriptores SIFT y LBP. En general, se considera una contribución el uso de PHOW en este contexto de aplicación analizado en la sección II B.

También, se encuentra el vocabulario visual generado. El vocabulario metálico propuesto, construido únicamente con características metálicas, difiere de los demás vocabularios propuestos en la literatura y se ha probado experimentalmente su superioridad con respecto a los vocabularios universales según los experimentos en [17]. Se considera, que es mejor que los vocabularios propuestos por Bastan y Mery en cuanto a su tamaño y descripción. Se ha mostrado que generalmente a mayor tamaño del vocabulario mejora la calidad de la clasificación [27]. Sin embargo, no se puede plantear lo mismo con el vocabulario usado por Turcsany, que no solo posee un mayor tamaño, sino que fue generado por otros dos vocabularios pertenecientes a cada clase de tamaño 600 cada uno. Se mostró en aquel estudio como este tipo de vocabularios puede lograr una mejor representación de los vectores de rasgos frente a los vocabularios universales.

El clasificador más utilizado en la mayoría de los trabajos que usan BoVW, debido a la forma de representar los datos mediante histogramas normalizados, es la máquina de soporte vectorial SVM. En este sentido, el trabajo de Mery difiere de los restantes. Se considera que la propuesta con el kernel $\chi^2$ para la SVM presenta mejor discriminación en el reconocimiento de armas en comparación con otros kernels. Este mejor comportamiento de este kernel se mostró también en los experimentos realizados en [17].

Teniendo en cuenta que en el trabajo de Mery y en este estudio se utilizaron imágenes de la misma base de datos GDXray se procede a realizar una comparación experimental. Los autores creen importante esta comparación debido a la actualidad y relevancia del trabajo de Mery.

En la tabla IV aparece la distribución de las series de imágenes usadas en la experimentación de ambos trabajos junto con la cantidad de imágenes por cada subgrupo. Las imágenes de la clase positiva se identifican con el signo '+' y las de la clase negativa con el signo '-'.

TABLA IV.
COMPARACIÓN DE LA EXPERIMENTACIÓN CON EL TRABAJO DE MERY

| Imágenes | | Este estudio | Mery |
|---|---|---|---|
| Entrenamiento | Serie | +B0049 y B0046 | +B0049 |
| | Cantidad | +12824 -12624 | +200 -700 |
| Validación | Serie | B0046 | +B0079 |
| | Cantidad | +5410 -5410 | +50 -300 |
| Prueba | Serie | B0044 | +B0079 |
| | Cantidad | +15947 -15947 | +99 -597 |

En el trabajo de Mery las imágenes de la clase negativa no arma fueron tomadas de otras series. Entrenamiento: B0050, B0051 y B0078, y para validación y prueba: B0080, B0081 y B0082.

---

[4] Local Binary Patterns
[5] Estos valores en toda la fila de la tabla son los tamaños de los vocabularios de cada estudio.
[6] El trabajo de Turcsany [7] utilizó la métrica Razón de Falsos Positivos (FPR) o razón de falsa alarma donde obtuvo 4,31%

Se puede observar en la tabla IV como la cantidad de imágenes en la experimentación usada en el trabajo de Mery en su propuesta de BoVW es pobre en comparación con la cantidad de imágenes usadas para ambas clases en el presente estudio. Esto se debe al uso del algoritmo de Búsqueda Selectiva para la generación de una mayor cantidad de combinaciones de imágenes para el protocolo de experimentación. Además, la serie de imágenes B0079 usada por Mery en los grupos de validación y prueba está formada por recortes manuales de las armas de las series B0046 y B0044 por lo que no añade información nueva a la clase positiva en comparación con este trabajo.

Debido a lo expuesto anteriormente, se considera la metodología de experimentación realizada mejor a la presentada por Mery (en la técnica que usó para BoVW) debido a la cantidad de imágenes en cada grupo que permitió un mejor entrenamiento y generalización del aprendizaje de la SVM. También se destacó en la sección III $E$ el procedimiento de la selección del parámetro regularizador $\lambda$ garantizando un resultado sin sobreajuste y de carácter generalizador. Todo lo expresado anteriormente explica, en parte, la diferencia en PPV y VPR de ambos trabajos visto en la tabla III. Además, los autores creen que la influencia del detector de puntos SIFT y el vocabulario visual de tamaño 300 como otros causantes en la diferencia de ambos resultados.

## V. Conclusiones

En este trabajo se desarrolló y evaluó un algoritmo de reconocimiento y detección automática de revólveres en imágenes de rayos X de una sola energía utilizando el modelo Bolsa de Palabras Visuales y el algoritmo Búsqueda Selectiva. La metodología propuesta fue probada en la base de datos GDXray [16] sobre las imágenes de las series B0044, B0046 y B0049 pertenecientes a equipajes con revólveres en diferentes posiciones y solapamiento con otros objetos. Las tres principales contribuciones de este trabajo son:

1) Se propuso la utilización del algoritmo de Búsqueda Selectiva para generar los recuadros imágenes de los grupos de entrenamiento, validación y prueba obteniéndose mayor cantidad de imágenes. Lo anterior, junto con el procedimiento de selección del parámetro de ajuste del regularizador posibilitó un mejor aprendizaje y generalización del clasificador.
2) La configuración del modelo Bolsa de Palabras Visuales (BoVW) para la generación de los histogramas de palabras visuales en cuanto al uso de PHOW como detector de puntos y el vocabulario metálico. Los autores consideran que es la primera vez que se utiliza BoVW con Búsqueda Selectiva en este contexto de aplicación.
3) Se comprobó la precisión del algoritmo de Búsqueda Selectiva en la detección de armas en imágenes de rayos X de equipaje de una sola energía, tanto con las anotaciones como junto con el clasificador. Para el conocimiento de los autores se cree que es la primera vez que se evalúa este algoritmo en este tipo de imágenes.

Los mejores resultados obtenidos en el reconocimiento del algoritmo fueron VPR = 92% y PPV = 80%. Esto constituye un 95,5% de exactitud general del algoritmo clasificador en el reconocimiento de revólveres. Es un resultado que compite con los resultados reportados en la literatura. Muestra la capacidad del algoritmo propuesto en el reconocimiento de la presencia de revólveres en este tipo de imágenes. Es un resultado mejor en comparación con la máxima sensibilidad reportada (90%) en los sistemas de inspección manual.

En cuanto al resultado de la detección se logró localizar un 37,5% de las armas de prueba acorde al criterio de calidad de solapamiento. Parece que el algoritmo de Búsqueda Selectiva presenta mejores resultados en la detección de objetos en imágenes de espectro visible según lo expuesto en [15]. En sentido general, el algoritmo propuesto es muy bueno reconociendo la presencia de las armas en las imágenes pero es pobre en su localización. Se cree que, se debe en parte, al proceso de fusión de los recuadros hipótesis de la Búsqueda Selectiva

Por el contrario, resultó significativa la evaluación del algoritmo de Búsqueda Selectiva en imágenes de rayos X de equipaje teniendo como referencia las anotaciones realizadas. Se obtuvo un resultado de 84% para un $k = 100$ y para un umbral de Pascal de 0.5 y un 98% para un umbral de Pascal de 0.4 con un promedio del índice de solapamiento de las mejores detecciones de 0.618.

Los autores consideran que la propuesta puede utilizarse como herramienta auxiliar en el proceso de inspección a pesar del tiempo de ejecución medido. Se realizó una prueba en una computadora Intel Core i5-4200U CPU @ 2.3GHz con una memoria RAM de 8GB. La ejecución total del algoritmo hasta la localización del arma en una imagen original de equipaje fue de promedio 41,5 segundos. Donde las etapas de mayor tiempo computacional fueron la ejecución del algoritmo de Búsqueda Selectiva con un promedio de 32,3 segundos y la extracción de características en toda la imagen con 5,7 segundos.

Esta prueba se realizó utilizando el MATLAB 2015b y sin realizarse modificaciones de eficiencia computacional al código original. Este tiempo es poco competitivo y se debe en gran medida a que se usaron las imágenes en sus dimensiones originales. Queda como recomendación la experimentación con las imágenes con su dimensión reducida.

A pesar de otros avances en las técnicas de visión computacional, el método BoVW continúa presentando desafíos y buenos resultados para la clasificación de objetos en imágenes y es un área en constante investigación. Este trabajo muestra la relevancia de la metodología propuesta con BoVW en el reconocimiento y detección de armas en imágenes de rayos X de una sola energía.

## Referencias

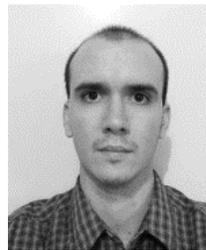

**David Castro Piñol** He obtained the BS degree in Telecommunication and Electronic Engineering from the Universidad de Oriente (UO), Cuba in 2015 with honors. He received a Master in Science of Management degree from the San Thomas University of Villanova, Florida, USA in 2018. He is working in UO since 2015, two years in the Study Center of Neurosciences, Image and Signals Processing researching in computer vision applications and now in the department of Telecommunications and Electronics. In addition, he is teaching undergraduate courses of digital signal processing and image processing. His current research interests include machine learning, and computer vision for multispectral satellite images analysis.

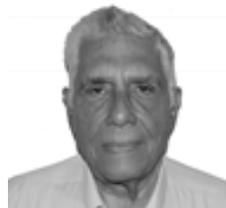

**Enrique J. Marañón Reyes** received the MD in electrical engineering in telecomunications in 1964 at the Universidad de la Habana, Havana, Cuba. Obtained the Ph.D. in Signals and Systems in 1985 at the Technical University of Praha. He is full professor at the department of Telecommunications and Electronics and Emeritus Professor of the Universidad de Oriente, Cuba. Is one of the founders 2004 and first Director of the Study Center of Neuroscience and Image and Signal Processing, Universidad de Oriente, Cuba. Where he currently works.